\documentclass[sigconf]{acmart}
\renewcommand\footnotetextcopyrightpermission[1]{} 
\usepackage{booktabs} 
\usepackage{epsfig}
\usepackage{endnotes}
\usepackage{enumerate}
\usepackage{graphicx}
\usepackage{epstopdf}
\usepackage{amsmath}
\usepackage{amssymb}
\usepackage{hyperref}
\usepackage{subfigure}
\usepackage{color}
\usepackage{paralist}
\usepackage{algorithm}
\usepackage{amsmath}
\usepackage{flushend}
\usepackage[strings]{underscore}

\usepackage{blindtext,graphicx}
\usepackage[absolute]{textpos}

\newcommand{\floor}[1]{\lfloor #1 \rfloor}
\newcommand{\ts}{\textsuperscript}

\begin{document}
\begin{textblock}{12}(2,0.2)
\noindent\small Published and presented in 33rd Annual Computer Security Applications Conference (ACSAC 2017)
\end{textblock}

\title{A Secure Mobile Authentication Alternative to Biometrics}
\author{Mozhgan Azimpourkivi}
\affiliation{
   \institution{Florida International University}
   \city{Miami}
   \state{FL}
   \country{USA}}
\author{Umut Topkara}
\affiliation{
   \institution{Bloomberg LP}
   \city{New york}
   \state{NY}
   \country{USA}}
\author{Bogdan Carbunar}
\affiliation{
   \institution{Florida International University}
   \city{Miami}
   \state{FL}
   \country{USA}}


\begin{abstract}

Biometrics are widely used for authentication in consumer devices and business
settings as they provide sufficiently strong security, instant verification and
convenience for users.  However, biometrics are hard to keep secret, stolen
biometrics pose lifelong security risks to users as they cannot be reset and
re-issued, and transactions authenticated by biometrics across different
systems are linkable and traceable back to the individual identity.  In
addition, their cost-benefit analysis does not include personal implications to
users, who are least prepared for the imminent negative outcomes, and are not
often given equally convenient alternative authentication options.

We introduce ai.lock, a secret image based authentication method for mobile
devices which uses an imaging sensor to reliably extract authentication
credentials similar to biometrics.  Despite lacking the regularities of
biometric image features, we show that ai.lock consistently extracts features
across authentication attempts from general user captured images, to
reconstruct credentials that can match and exceed the security of biometrics
(EER = 0.71\%). ai.lock only stores a ``hash'' of the object's image. We
measure the security of ai.lock against brute force attacks on more than 3.5
billion authentication instances built from more than 250,000 images of real
objects, and 100,000 synthetically generated images using a generative
adversarial network trained on object images.  We show that the ai.lock Shannon
entropy is superior to a fingerprint based authentication built into popular
mobile devices. 
\end{abstract}


\maketitle
\thispagestyle{empty}

\section{Introduction}

Existing solutions to the complex mobile authentication equation have
significant problems. For instance, while biometric authentication provides
sufficiently strong security, instant verification and convenience for users,
biometrics are also hard to keep secret and pose lifelong security risks to
users when stolen, as they cannot be reset and re-issued, More importantly, as
surrendering biometrics may become de facto
mandatory~\cite{Aadhaar,keenanhidden}, existing
vulnerabilities~\cite{vrfaceattack,facebookface,ACJP14,GRGBFOG12}, coupled with
the compromise of large scale biometrics databases~\cite{OPM}, raise
significant long term security concerns, especially as transactions
authenticated by biometrics across different systems are linkable and traceable
back to the individual identity. Further, token-based authentication solutions,
e.g., SecurID~\cite{SecurID}, require an expensive
infrastructure~\cite{SecurIDCost} (e.g. for issuing, managing, synchronizing
the token).



A secret image based authentication approach, where users authenticate using
arbitrary images they capture with the device camera, may address several of
the above problems. For instance, the authentication is not tied to a visual of
the user's body, but that of a personal accessory, object, or scene. As
illustrated in Figure~\ref{fig:system}, a user sets her {\it reference}
credential to be an image of a nearby object or scene. To authenticate, the
user captures a {\it candidate} image; the authentication succeeds only if the
candidate image contains the same object or scene as the reference image. This
improves on (1) biometrics, by freeing users from personal harm, providing
plausible deniability, allowing multiple keys, and making revocation and change
of secret simple and (2) token-based authentication, by eliminating the need
for an expensive infrastructure. Visual token-based solutions (e.g., based on
barcodes or QR codes)~\cite{sib,webticket} can be seen as special cases of
secret image based authentication.

However, this approach raises new challenges. First, an adversary who
captures or compromises the device that stores the user's reference credentials
(e.g. mobile device, remote server) and has access to its storage, should not
be able to learn information about the reference credentials or their features.
Second, while biometric features such as ridge flow of fingerprints or eye
socket contours of faces, can be captured with engineered features and are
invariant for a given user, images of objects and general scenes lack a well
defined set of features that can be accurately used for authentication
purposes.  Improper features will generate (i) high false accept rates (FAR),
e.g., due to non-similar images with similar feature values, and (ii) high
false reject rates (FRR) that occur due to angle, distance and illumination
changes between the capture circumstances of reference and candidate images.

In a first contribution, we introduce {\it ai.lock}, a practical, secure and
efficient image based authentication system that converts general mobile device
captured images into biometric-like structures, to be used in conjunction with
secure sketch constructs and provide secure authentication and storage of
credentials [$\S$~\ref{sec:ai.lock}].

To extract invariant features for image based authentication, ai.lock leverages
(1) the ability of Deep Neural Networks (DNNs) to learn representations of the
input space (i.e., {\it embedding vectors} of images) that reflect the salient
underlying explanatory factors of the data, (2) Principal Component Analysis
(PCA)~\cite{P01} to identify more distinguishing components of
the embedding vectors and (3) Locality Sensitive Hashing (LSH) to map the
resulting components to binary space, while preserving similarity properties in
the input space. We call the resulting binary values {\it imageprints}. ai.lock
builds on a secure sketch variant~\cite{DORS08} to securely store reference
imageprints and match them to candidate imageprints.

In a second contribution, we propose the LSH-inspired notion of {\it locality
sensitive image mapping} functions ($\delta$-LSIM), that convert images to
binary strings that preserve the ``similarity'' relationships of the input
space, for a desired similarity definition [$\S$~\ref{sec:problem}]. A
$\delta$-LSIM function can be used to efficiently match images based on their
extracted binary imageprints.

\begin{table}
\centering
\footnotesize{
\resizebox{0.47\textwidth}{!}{%
\textsf{
\begin{tabular}{l c c c c}
\toprule
\textbf{} & \textbf{FAR} & \textbf{EER} & \textbf{Shannon Entropy} & \textbf{Dataset}\\
\textbf{Solution} & \textbf{(\%)} & \textbf{(\%)} & \textbf{(bits)} & \textbf{size}\\
\midrule
ai.lock (MLMS) & 0.0004 & - & 18.02 & 2 $\times 10^9$\\
ai.lock (MLSS) & 0.0015 & 0.17 & 16.02 & 6 $\times 10^6$\\
\midrule
iPhone TouchID~\cite{tochidentropy} & 0.0020 & - & 15.61 & -\\
\midrule
Deepface~\cite{deepface} (face) & - & 8.6 & - & $>$ 0.5 $\times 10^9$ \\
SoundProof~\cite{soundproof} (sound) & 0.1 & 0.2 & 9.97 & $>$ 2 $\times 10^6$ \\
~\cite{eyemovement} (eye movement) & 0.06 & 6.2 & 10.70 & $1,602$\\
\midrule
RSA SecurID~\cite{SecurID} & - & - & 19.93 & - \\
\midrule
Text-based password ~\cite{yahoopasswords} & - & - & 10-20 & 7 $\times 10^7$ \\
\bottomrule
\end{tabular}}}}
\caption{ai.lock variants vs. commercial and academic biometric, token-based 
authentication solutions, and text passwords. ai.lock MLSS variant has no false 
rejects, as it is evaluated under attack samples only. {\bf Under large scale 
datasets of powerful attacks, ai.lock achieves better entropy than state-of-the-art
biometric solutions}.}
\label{table:comparison}
\vspace{-25pt}
\end{table}

Further, we develop brute force image based attacks that aim to defeat ai.lock.
First, we perform {\it real image attacks}, that use manually collected and publicly
available image datasets. To evaluate ai.lock on large scale attack
images, we develop {\it synthetic image attacks} that use images produced by
generative models~\cite{DCGAN}. To evaluate the resilience of stored
credentials, we introduce {\it synthetic credential attacks}, that use
authentication credentials generated with the same distribution of the credentials
extracted from manually collected images [$\S$~\ref{sec:model:adversary}].  We
have captured, collected and generated datasets of 250,332 images, and
generated 1 million synthetic credentials [$\S$~\ref{sec:data}].  We have used
these datasets to generate attack datasets containing more than 3.5 billion
(3,567,458,830) authentication instances [$\S$~\ref{sec:eval:data}].


We have implemented an ai.lock in Android using Tensorflow~\cite{tensorflowpaper} 
and show that it is resilient to attacks. Its FAR on 140 million synthetic image 
attack samples is $0.2 \times 10^{-6}\%$. ai.lock was unbreakable when tested 
with 1.4 billion synthetic credential attack samples. The estimated Shannon entropy
~\cite{shannonEntropy} of ai.lock on 2 billion image pairs is 18.02 bits,
comparing favorably with state-of-the-art biometric solutions (see
Table~\ref{table:comparison}). Further, we show that ai.lock is a $\delta$-LSIM
function, over images that we collected [$\S$~\ref{sec:eval:lsim}]. ai.lock is
fast, imposing an overhead of under 1s on a Nexus 6P device. 
We have released the code and data on 
https://github.com/casprlab/ai.lock.

\section{Model and Applications}
\label{sec:model}

\begin{figure}[t!]
\centering   
\includegraphics[width=0.47\textwidth,keepaspectratio]{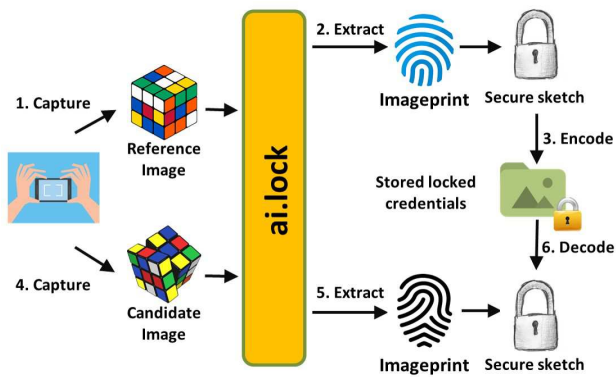}
\vspace{-10pt}
\caption{
{\bf ai.lock model and scenario}. The user captures the image of an object or
scene with the device camera. ai.lock converts the image to a binary {\it
imageprint}, and uses it as a biometric, in conjunction with a secure sketch
solution, to securely store authentication information on the device 
or on a remote server.
The user can authenticate only if she is able to capture another image of
the same object or scene.}
\label{fig:system}
\vspace{-15pt}
\end{figure}

We consider a user that has a camera equipped device, e.g., smartphone or
tablet, a resource constrained device such as a smart watch/glasses, or a complex
cyber-physical system such as a car. The user needs to authenticate to the
device or an application back-end, or authenticate through the device to a
remote service. For this, we assume that the user can select and easily access
a physical object or scene. To set her password, the user captures the image of 
an object/scene with the device camera, see Figure~\ref{fig:system} for an 
illustration. ai.lock extracts a set of features from the user's captured 
{\it reference} image, then stores this information (imageprint) securely  
either on the device or on a remote server. We note that, in the former case, the 
device can associate the reference image with the user's authentication credentials 
(e.g. OAuth~\cite{rfc6749}) for multiple remote services. 
To authenticate, the user needs to capture another image. The user is able to
authenticate only if the {\it candidate} image is of the same object or scene
as the reference image. Similar to e.g., text passwords, the user can choose to
reuse objects across multiple services, or use a unique object per service.
Using a unique object per service will affect memorability.  However, due to
the image superiority effect~\cite{pictorialSuperiority}, objects may be easier
to remember than text passwords. In the following, we describe a few
applications of this model.

\noindent
{\bf Alternative to biometric authentication}.
Instead of authenticating with her sensitive and non-replaceable biometrics
(face, fingerprint), the user uses a unique nearby scene or object that she
carries, e.g., a trinket, Rubik's cube with a unique pattern, printed random
art, etc.  ai.lock moves the source of information from the user to an
externality, as it does not require a visual of the user's body, but that of a
personal accessory, object, or scene that the user can recreate at
authentication time.  ai.lock improves on biometrics by freeing users from
personal harm, providing plausible deniability, allowing multiple keys, and
making revocation and change of secret simple.

\noindent
{\bf Location based authentication}.
The user chooses as password an image of a unique scene at a frequented
location (office, home, coffee shop), e.g., section of book shelf, painting,
desk clutter. This approach can be generalized to 
enable location based access control, e.g., to
provide restricted access to files and networks in less secure locations.

\noindent
{\bf Cyber-physical system authentication}.
Our model supports authentication to cyber-physical systems, including car and
door locks, thermostat and alarm systems, where key and PIN entry
hardware~\cite{schlage,securiCode} is replaced with a camera. To authenticate,
the user needs to present her unique but replaceable authentication object to
the camera.


\subsection{Adversary Model}
\label{sec:model:adversary}

We assume an active adversary who can physically capture or compromise the
device that stores the user credentials. Such an adversary can not only access
the stored credentials, but also any keying material stored on the device, then
use it to recover encrypted data and use it to authenticate through the proper
channels.  However, we assume that the adversary does not have control over the
authentication device while the user authenticates (e.g., by installing malware).
We also assume an adversary
with {\it incomplete surveillance}~\cite{Filardo2012}, i.e., who can physically
observe the victim during authentication but cannot capture the details of the
secret object.

Furthermore, we assume that the adversary has ``blackbox access'' to the
authentication solution, thus can efficiently feed it images of his choice and
capture the corresponding imageprint.  The adversary can use
this output to learn information from the stored credentials. 
More specifically, we consider the following attacks:

$\bullet$
{\bf Real image attack}.
The adversary collects large datasets of images, e.g., manually using a mobile
camera, and online.  Then, in a brute force approach, he matches each image as
an authentication instance against the stored reference credentials until
success.

$\bullet$
{\bf Synthetic image attack}.
The adversary uses the previously collected images to train a generative model,
e.g.~\cite{GAN}, that captures essential traits of the images, then uses the
trained model to generate a large dataset of synthetic images. Finally, the
adversary matches each such image against the reference credentials.

$\bullet$
{\bf Synthetic credential attack}.
Instead of images, the adversary queries the authentication system with binary
imageprints. For this, the adversary extracts the imageprints generated by 
the authentication solution on real images of his choice. 
He then generates a large dataset of synthetic credentials that follow
the same distribution as the extracted credentials.  Finally, he matches each
synthetic credential exhaustively against the reference credentials.

$\bullet$
{\bf Object/scene guessing attack}.
While we do not consider shoulder surfing attacks which also apply to face
based authentication~\cite{vrfaceattack,livenessface}, we assume an adversary
that is able to guess the victim's secret object/scene type. The adversary then
collects a dataset of images containing the same object or scene type, then
uses them to brute force ai.lock (see Appendix~\ref{sec:appendix:shouldersurfing}).



Finally, we assume the use of standard secure communication channels for the
remote authentication scenario where the user credentials are stored on a
server.

\section{Problem Definition}
\label{sec:problem}

Let $\mathbb{I}$ denote the space of images that can be captured by a user with
a camera. Let $sim: \mathbb{I} \times \mathbb{I} \rightarrow \{0,1\}$ be a
function that returns true when its input images have been taken with the same
camera and are of the same object or scene, and false otherwise.

Informally, the {\it image based authentication} problem seeks to identify a
{\it store} function $S : \mathbb{I} \rightarrow \{0,1\}^k$, and an {\it
authentication} function $Auth : \{0,1\}^k \times \{0,1\}^* \rightarrow
\{0,1\}$ (for a parameter $k$) that satisfy the following properties.  First,
it is hard for any adversary with access to only $S(R)$, for a reference image
$R \in \mathbb{I}$, to learn information about $R$. That is, $S$ imposes a
small entropy reduction on its input image. Second, for any candidate string $C
\in \{0,1\}^*$, $Auth(S(R), C) = 1$ only if $C \in \mathbb{I}$ and $sim(R, C) =
1$.  Thus, a candidate input to the $Auth$ function succeeds only if it is a
camera captured image of the same object or scene as the reference image.

We observe that the {\it secure sketch} of~\cite{DORS08} solves this problem
for biometrics: given a biometric input, the secure sketch outputs a value that
reveals little about the input, but allows its reconstruction from another
biometric input that is ``similar''. Therefore, the image based authentication
problem can be reduced to the problem of transforming camera captured images of 
arbitrary objects and scenes into biometric-like structures.

Hence, we introduce the LSH-related notion of {\it locality sensitive
image mapping} functions. Specifically, let $d: \{0,1\}^{\lambda} \times
\{0,1\}^{\lambda} \rightarrow \mathbb{R}$ be a distance function (e.g.,
Hamming), where $\lambda$ is a system parameter. Then, for a given $\delta \in
[0,1]$, a $\delta$-Locality Sensitive Image Mapping (LSIM) function $h$
satisfies the following properties:

\begin{definition}
$h: \mathbb{I} \rightarrow \{0,1\}^\lambda$ is a $\delta$-LSIM
function if there exist probabilities $P_1$ and $P_2$, $P_1 > P_2$, s.t.:

\begin{enumerate}

\item
For any two images $I_1, I_2 \in \mathbb{I}$, if $sim(I_1, I_2) = true$, then
$\frac{d(h(I_1),h(I_2))}{\lambda} < \delta$ with probability $P_1$.

\item
For any two images $I_1, I_2 \in \mathbb{I}$, if $sim(I_1, I_2) = false$,
then $\frac{d(h(I_1),h(I_2))}{\lambda} > \delta$ with probability $P_2$.

\end{enumerate}
\label{def:dlsim}
\end{definition}


\section{Background \& Related Work}
\label{sec:background}

To build ai.lock we leverage deep learning based feature extraction, locality
sensitive hashing and secure sketch constructs.  In the following, we briefly
describe these concepts.

\subsection{Biometric Protection}
\label{sec:background:ss}


Our work is related to the problem of protecting biometric templates. We
summarize biometric protection solutions, that can be classified into fuzzy
biometric protection and feature transformation
approaches~\cite{surveybiometric}.

\noindent
{\bf Fuzzy biometric template protection}.
This approach leverages error correcting codes to verify biometric data.
Techniques include secure sketch and fuzzy extractor~\cite{DORS08}, fuzzy
vault~\cite{JS06} and fuzzy commitment~\cite{JW99}, and have been applied to
different biometric data, e.g. palm and hand~\cite{lalithamani2015palm}. 

In this paper, we extend the secure sketch under the Hamming distance solution
from~\cite{DORS08}: reconstruct the biometric credential, then compare its hash
against a stored value. We briefly describe here the password set and
authentication procedures that we use based on ai.lock generated imageprints 
(see $\S$~\ref{sec:ai.lock}). Let $ECC$ be a binary error correcting
code, with the corresponding decoding function $D$, and let $H$ be a
cryptographic hash function.

$\bullet$
{\bf Image password set}.
Let $R$ be the reference image captured by the user and let $\pi_R = \pi(R)$ be
its ai.lock computed imageprint. Generate a random vector $x$, then compute and
store the authentication credentials, $SS(R,x) = \langle SS_1, SS_2 \rangle$,
where $SS_1 = \pi_R \oplus ECC(x)$ and $SS_2 = H(x)$.

$\bullet$
{\bf Images based authentication}.
Let $C$ be the user captured candidate image, and let $\pi_C = \pi(C)$ be its
ai.lock computed imageprint ($\S$~\ref{sec:ai.lock}). Retrieve the stored $SS$
value and compute $x' = D(\pi_C \oplus SS_1)$. The authentication succeeds if
$H(x') = SS_2$.


\noindent
{\bf Transformation based biometric template protection}.
A transformation is applied both to the biometric template and the biometric
candidate, and the matching process is performed on the transformed data. 
In an invertible transformation (a.k.a., {\it salting}~\cite{surveybiometric}), 
a key, e.g., a password, is used as a parameter to define the transformation 
function ~\cite{biohashing}. 
The security of this approach depends on the ability to protect the key. 
In contrast, in non-invertible schemes ~\cite{maiorana2010cancelable,cancelablebiometrics} 
a one-way transformation functions is used to protect the biometric template, making 
the inversion of a transformed template computationally hard even when the key is revealed.

\noindent
{\bf Hybrid approaches}.
Hybrid transformation and fuzzy protection approaches have also been proposed.
Nandakumar et al.~\cite{nandakumar2007hardening} introduced an approach to make 
the fingerprint fuzzy value stronger using a password as salt. Song et al.
~\cite{ong2008application} used discrete hashing to transform the fingerprint 
biometric, which is then encoded and verified using error correcting codes.

\subsection{Deep Neural Networks (DNNs)}
\label{sec:background:dnn}

%
%
%
%

Empirical results have shown the effectiveness of representations learned by
DNNs for image classification~\cite{transferable, donahue2014decaf, 
multimediaclassification}, and for
the verification of different biometric information~\cite{siamese, menotti,
FSDS}. However, ai.lock differs in its need to ensure that two object images
contain the exact same object, for the purpose of authentication.
%
%
%
%
%
ai.lock exploits the ability of DNNs to learn features of the input image space
that capture the important underlying explanatory factors. We conjecture that
these features will have small variations among images of the same object or
scene, captured in different circumstances. 


\noindent
{\bf Pretrained Inception.v3}.
Training a DNN with millions of parameters is computationally expensive and
requires a large training dataset of labeled data, rarely available in
practice. Instead, we employed a {\it transfer learning}~\cite{shao2015transfer}
approach: obtain a trained DNN and use it for a similar
task.  For image feature extraction, we use Inception.v3
~\cite{inception3} network pretrained on ImageNet dataset \cite{imagenet}, of
1.2 million images of 1,000 different object categories, for image classification. 



\subsection{Locality Sensitive Hashing}
\label{sec:background:lsh}

Locality Sensitive Hashing (LSH) seeks to reduce the dimensionality of data,
while probabilistically preserving the distance properties of the input space.
It was initially used to solve the near neighbor search problem in high
dimensional spaces~\cite{IM98}. 
While seemingly the ideal candidate to provide
the ai.lock functionality, LSH does not work well on images: images of the same
scene or object, captured in different conditions, e.g., angle,
distance, illumination, will have dramatically different pixel values, leading
to a high distance between the images and thus also between their LSH values.

We use however Charikar's~\cite{C02} LSH as a building block in ai.lock.
Charikar's~\cite{C02} LSH defines a family of hash functions in the space
$\mathcal{R}^d$. Specifically, the LSH function $h_r$ is based on a randomly
chosen $d$-dimensional Gaussian vector with independent components $r \in
\mathcal{R}^d$, where $h_r(u) = 1$ if $r \cdot u \ge 0$ and $h_r(u) = 0$ if $r
\cdot u < 0$, where $\cdot$ denotes the inner product. This function provides
the property that $Pr[h_r(u) = h_r(v)] = 1 - \frac{\theta(u,v)}{\pi}$, for any
vectors $u$ and $v$, where $\theta(u,v)$ denotes the angle between the input
vectors.

\subsection{Privacy Preserving Image Matching}

\begin{figure}
\centering 
\includegraphics[width=0.49\textwidth, keepaspectratio]{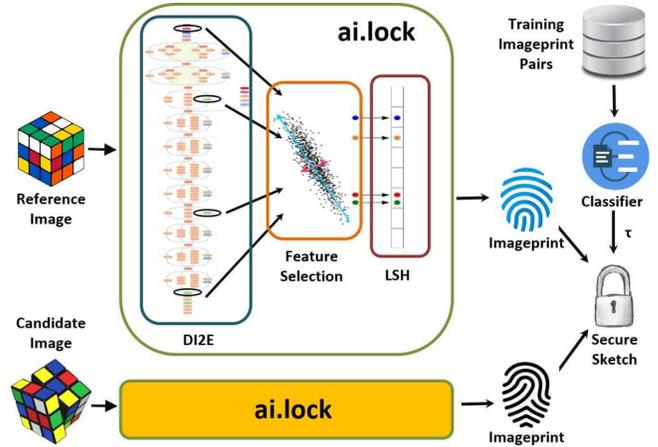}
\caption{{\bf ai.lock architecture}. ai.lock processes the input image through
a deep neural network (i.e., Inception.v3),
selects relevant features, then
uses locality sensitive hashing to map them to a binary imageprint. ai.lock
uses a classifier to identify the ideal {\it error tolerance threshold}
($\tau$), used by the secure sketch block to lock and match imageprints.}
\label{fig:ai.lock}
\vspace{-15pt}
\end{figure}

Traditional approaches to object matching and object recognition, e.g.,
SIFT~\cite{sift} and SURF~\cite{surf}, rely on extracting identifying features
or (robust/invariant) keypoints and their descriptors at specific
locations on the image.  Several solutions have been proposed for the secure
image matching problem that could be applied to the image based authentication
task.  SecSIFT~\cite{secsift} employed order-preserving encryption and
distributes the SIFT computation among different servers. Hsu et
al.~\cite{privatesift} proposed a privacy preserving SIFT based on homomorphic
encryption, while Bai et al.~\cite{encsurf} performed SURF feature extraction in
encrypted domain using Paillier's homomorphic cryptosystem. Wang et
al.~\cite{wang2016securesurf} improve the SURF algorithm in 
encrypted domain by designing protocols for secure multiplication and
comparison, that employ a ``somewhat'' homomorphic encryption.
These approaches are not practical on mobile devices, due to the high cost of
homomorphic encryption and the large number of keypoints (up to thousands per
image). 


\subsection{Token-Based Authentication}
\label{sec:background:token}

In previous work~\cite{ATC17} we have evaluated the usability of Pixie, a 
trinket based authentication solution that employs slightly outdated image
processing techniques to extract features (i.e., ``keypoints'') and match
user captured images. Pixie has an important drawback when deployed on mobile
devices: the image keypoints that it extracts need to be stored and matched in 
cleartext on vulnerable devices. In contrast, ai.lock uses state of the art, 
deep neural network based image feature extraction along with LSH to extract
binary imageprints that are robust to changes in image capture conditions.
The imageprints can be securely stored and matched using
secure sketches. This makes ai.lock resilient to device capture attacks.
Furthermore, on larger and more complex attack datasets, the use of DNNs
enabled ai.lock to achieve false accept rates that are at least 2 orders of
magnitude smaller than those of Pixie ($\le 0.0015\%$ vs. $0.2-0.8\%$),
for similar FRRs (4\%).

ai.lock's secret physical object is similar to token-based authentication,
either hardware or software. For instance, SecurID~\cite{SecurID} generates
pseudo-random, 6 digit authentication codes. ai.lock's Shannon entropy is
slightly lower than SecurID's 19.93 bits (see Table~\ref{table:comparison} for
comparison). Several authentication solutions use visual tokens (e.g., barcodes
or QR codes). For instance, McCune, et al.~\cite{sib} proposed
Seeing-is-Believing, a schema that relies on a visual authentication channel
that is realized through scanning a barcode.  Hayashi et al.~\cite{webticket}
introduced WebTicket, a web account management system that asks the user to
print or store a 2D barcode on a secondary device and present it to the
authentication device's webcam in order to authenticate to a remote service.
Token-based authentication requires an expensive
infrastructure~\cite{SecurIDCost} (e.g. for issuing, managing, synchronizing
the token). ai.lock provides a ``hash-like" construct for arbitrary object
images, making objects usable as passwords, with the existing infrastructure.


Other approaches exist that seek to transform biometrics into tokens that the
user needs to carry, with important implications on biometric privacy and
revocation capabilities. For instance, TAPS~\cite{TAPS} is a glove sticker with
a unique fingerprint intended for TouchID.

\section{The ai.lock Solution}
\label{sec:ai.lock}

We introduce ai.lock, the first locality sensitive image mapping function,
and a practical image based authentication system. In the
following, we describe the basic solution, then introduce two performance
enhancing extensions.

\subsection{ai.lock: The Basic (SLSS) Solution}
\label{sec:ai.lock:slss}

ai.lock consists of 3 main modules (see Figure~\ref{fig:ai.lock}): (1) 
deep image-to-embedding (DI2E) conversion module (2) feature selection module, and (3)
LSH based binary mapping module. We now describe each module and its interface
with the secure sketch module (see $\S$~\ref{sec:background:ss}).
Table~\ref{table:param} summarizes the important ai.lock parameter notations.

\begin{table}
\centering
\resizebox{0.47\textwidth}{!}{%
\textsf{
\begin{tabular}{c | l}
\toprule
\textbf{Symbol} & \textbf{Description}\\
\midrule
$\lambda$ & Length of the imageprint for a single image segment\\
$\tau$     & Error tolerance threshold\\
$c$             & Correctable number of bits \\
$s$       & Number of image segments in multi segment schema\\
$t$       & \multicolumn{1}{p{7cm}}{\raggedright Segment-based secret sharing threshold} \\
\bottomrule
\end{tabular}
}
}
\caption{ai.lock notations.}
\label{table:param}
\vspace{-15pt}
\end{table}

\noindent
{\bf Deep image to embedding (DI2E) module}.
Let $I$ be the fixed size input image. Let $Emb: \mathcal{I} \rightarrow
\mathcal{R}^e$ be a function that converts images into feature vectors 
of size $e$. We call $Emb(I)$ the {\it embedding vector}, an
abstract representation of $I$.
%
%
To extract $Emb(I)$, ai.lock uses the activations of a certain layer of  
Inception.v3 DNN~\cite{inception3} when $I$ is the input to the network. 
Let $e$ denote the size of the output of the layer of the DNN used by ai.lock. 
Thus, $Emb(I) \in \mathcal{R}^e$.


\noindent
{\bf Feature selection module}.
We have observed that not all the components in the embedding feature vectors
are relevant to our task (see $\S$~\ref{sec:eval:parameters}). Therefore, we 
reduce the dimensionality of the feature vectors to improve the performance 
and decrease the processing burden of ai.lock. 
Let $P: \mathcal{R}^e \rightarrow \mathcal{R}^p$, where $p < e$ be a
function that reduces the features of an embedding to the ones that are most
important. ai.lock uses PCA with component 
range selection as the $P$ function, and applies it to $Emb(I)$ to find a set of 
components that can reflect the distinguishing features of images. 
Thus, the vector produced by feature selection module is $P(Emb(I)) \in \mathcal{R}^p$.

\noindent
{\bf LSH based binary mapping module}.
In a third step, ai.lock seeks to map $P(Emb(I))$ to a binary space of size
$\lambda$ that preserves the similarity properties of the input space. 
%
%
%
To address this problem, we use the LSH 
scheme proposed by Charikar ~\cite{C02}. Let $L: \mathcal{R}^p
\rightarrow \{0,1\}^\lambda$ be such a mapping function. ai.lock uses as $L$, a
random binary projection LSH as follows. Let $M$ be a matrix of size $p \times
\lambda$, i.e. $\lambda$ randomly chosen $p$-dimensional Gaussian vectors with
independent components. Calculate $b$ as the dot product of $P(Emb(I))$ and
$M$.
%
%
For each coordinate of $b$, output either 0 or 1, based on
the sign of the value of the coordinate. We call this binary representation of
the input image $I$, i.e. $\pi(I) = L(P(Emb(I)))$, its {\it imageprint}. We
denote the length of a single imageprint by $\lambda$. Note that, the hash value
for the Charikar's method is a single bit ($\lambda$ = 1). Therefore, $L$ can
be viewed as a function that returns a concatenation of $\lambda$ such random
projection bits. 

In $\S$~\ref{sec:eval:lsim} we provide empirical evidence that the function $h =
L \circ P \circ Emb$ is a $(\tau)$-LSIM transform 
(see $\S$~\ref{sec:eval:parameters}), for specific $\tau$ values.

\noindent
{\bf Secure sketch}.
ai.lock extends the secure sketch under the Hamming distance of~\cite{DORS08}
to securely store the binary imageprint credentials and perform image password
set and image based authentication as described in
$\S$~\ref{sec:background:ss}.

In the following, we introduce two ai.lock extensions, intended to increase the
entropy provided by ai.lock's output. First, we modify ai.lock to use the
embedding vectors obtained from multiple layers of Inception.v3 network.
Second, we extend ai.lock to split the input image into multiple overlapping
segments and concatenate their resulting binary representations.


\subsection{ai.lock with Multiple DNN Layers}
\label{sec:ai.lock:ml}

Representations learned by a DNN are distributed in different layers of these
networks. The lower (initial) layers of convolutional neural networks learn 
low level filters (e.g. lines, edges), while deeper layers learn more abstract 
representations ~\cite{convnetvisualization}. The use of a single
DNN layer prevents the basic ai.lock solution from taking advantage of both 
filters.

To address this issue, we propose an ai.lock extension that collects the
embedding vectors from multiple ($l$) layers of Inception.v3 network. In
addition, we modify the basic ai.lock feature extractor module as follows. The
Principal Components (PCs) of activations for each layer are computed
separately and are mapped to a separate binary string of length $\lambda$.
Then, the binary strings constructed from different layers are concatenated to
create a single imageprint for the input image. Thus, the length of the
imageprint increases linearly with the number of layers used in this schema.

\subsection{ai.lock with Multiple Image Segments}
\label{sec:ai.lock:multiseg}

\begin{figure}[t!]
\centering
\includegraphics[width=0.40\textwidth, height=1.5in]{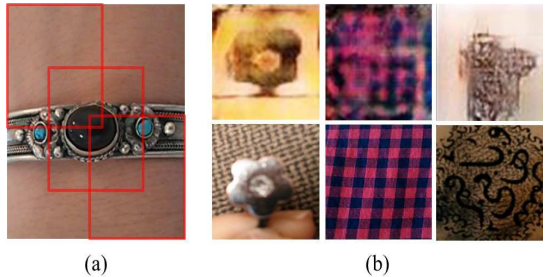}
\caption{
(a) 3 overlapping segments of an image.
(b) Top: sample images generated by DCGAN, Bottom: visually similar
images in Nexus Dataset to images generated by DCGAN.}
\label{fig:trinket}
\vspace{-15pt}
\end{figure}

We divide the original image into $s$ overlapping segments (see
Figure~\ref{fig:trinket}(a)). We then run the basic ai.lock
over each segment separately to produce $s$ different imageprints of length
$\lambda$. However, we identify the PCs for the embedding vectors of each
segment based on the whole size images. The intuition for this choice is that
random image segments are not good samples of real objects and may confuse the
PCA. We then generate the imageprint of the original, whole size image, as the
concatenation of the imageprints of its segments.

\noindent
{\bf Secure sketch sharing}.
We extend the secure sketch solution with a $(t, s)$-secret sharing scheme.
Specifically, let $x_1,..,x_s$ be $(t, s)$-shares of the random $x$, i.e.,
given any $t$ shares, one can reconstruct $x$.  Given a reference image $R$,
let $R^{(1)},..R^{(s)}$ be its segments, and let $\pi_R^{(i)} = \pi(R^{(i)})$,
$i=1..s$ be their imageprints.  Then, we store $SS(R,x) = \langle
SS_1^{(1)},..,SS_1^{(s)}, SS_2 \rangle$, where $SS_1^{(i)} = \pi_R^{(i)} \oplus
ECC(x_i)$ and $SS_2 = H(x)$. To authenticate, the user needs to provide a
candidate image $C$, whose segments $C^{(i)}, i=1..s$ produce imageprints
$\pi_C^{(i)} = \pi(C^{(i)})$ that are able to recover 
at least $t$ of $x$'s shares $x_i$.


\section{Implementation \& Data}

We build ai.lock on top of the Tensorflow implementation for 
Inception.v3 network \cite{InceptionTensorflow}. For the error correcting code
of secure sketches, we use a BCH~\cite{H59,BC60} open source
library~\cite{bchlib}, for syndrome computation and syndrome
decoding with correction capacity of up to $c$ bits.  The value for $c$ is
calculated empirically using the training dataset (see $\S$~\ref{sec:eval:parameters})

\noindent
{\bf Basic (SLSS) ai.lock}.
In the basic ai.lock solution, we use the output of the last hidden layer of
Inception.v3 network, before the softmax classifier, consisting of $2,048$ float
values. Our intuition is that this layer provides a compact representation (set
of features) for the input image objects, that is efficiently separable by the
softmax classifier.

\noindent
{\bf Multi layer ai.lock}.
For the multi DNN layer ai.lock variants, we have used 2 layers ($l = 2$). The
first layer is the ``Mixed{\_}8/Pool{\_}0'' layer and the second layer is the
last hidden layer in Inception.v3.  The embedding vector for the
``Mixed{\_}8/Pool{\_}0'' consists of $49,152$ float values.  As described in
$\S$~\ref{sec:ai.lock:multiseg}, the embedding vectors of each layer are
separately processed by the feature selection and LSH modules; the resulting
binary strings are concatenated to form the imageprint of size $2 \lambda$.


\noindent
{\bf Multi segment ai.lock}.
For the multi segment ai.lock variant, we split the image into multiple
segments that we process independently. Particularly, we consider 5 overlapping 
segments, cropped from the top-left, bottom-left, top-right, bottom-right and the 
center of the image.  We generate segments whose width and height is equal
to the width and height of the initial image divided by 2, plus 50 pixels to
ensure overlap.  The extra 50 pixels are added to the interior sides for the
side segments. For the middle segment, 25 pixels are added to each of its
sides. Each segment is then independently processed with the
basic ai.lock (i.e., last hidden layer of Inception.v3, PCA, LSH).

\noindent
{\bf Multi layer multi segment ai.lock}.
This is a hybrid of the above variants: split the image into 5 overlapping
parts, then process each part through Inception.v3 network, and extract the
activation vectors for each of the two layers of Inception.v3 (the last
hidden layer and Mixed{\_}8/Pool{\_}0 layer). The output of each layer for each
segment is separately processed as in the basic ai.lock. Thus, the resulting
imageprint of the image has $10 \lambda$ bits.

\subsection{Primary Data Sets}
\label{sec:data}

\subsubsection{Real Images}
\label{sec:data:real}
\hfill \break
\noindent
{\bf Nexus dataset}.
We have used a Nexus 4 device to capture 1,400 photos of 350 objects, belonging
to 33 object categories. Example of object categories in this dataset includes
watches, shoes, jewelry, shirt patterns, and credit cards. We have captured 4
images of each object, that differ in background and lighting conditions.

\noindent
{\bf ALOI dataset}.
We have used the ``illumination direction'' subset of the Amsterdam Library of Object
Images (ALOI)~\cite{aloi} dataset. This dataset includes 24 different images of
1000 unique objects (24,000 in total) that are taken under different illumination
angles.

\noindent
{\bf Google dataset}.
We have used Google's image search to retrieve at least 200 images from each of
the 33 object categories of the Nexus image dataset, for a total of 7,853
images. This dataset forms the basis of a ``targeted'' attack. 

\noindent
{\bf YFCC100M toy dataset}.
We have extracted a subset of the Yahoo Flickr Creative Commons 100M
(YFCC100M)~\cite{yahooflickr} image dataset (100 million Flickr images),
of 126,600 Flickr images tagged with the ``toy'' keyword, and not with
``human'' or ``animal'' keywords.

\subsubsection{Synthetic Data}
\label{sec:data:synthetic}
\hfill \break
\noindent
{\bf Synthetic image dataset}.
Manually capturing the Nexus dataset was a difficult and time consuming 
process. In order to efficiently generate a large dataset of similar images, 
we have leveraged the ability of generative models to discover an abstract 
representation that captures the essence of the training samples. Generative
models, including 
%
%
Generative Adversarial Networks (GAN)~\cite{GAN}, are trained to generate samples that are
similar to the data they have been trained on. Such models have been shown to
be suitable for representation learning tasks, e.g.,~\cite{DCGAN}.

We have used a DCGAN network~\cite{DCGAN} to generate a large set of synthetic
images that are similar to the images in the Nexus dataset.  Specifically, we
have trained a DCGAN network~\cite{DCGAN} using the images of the Nexus dataset
for 100 training epochs. Image augmentation, e.g., rotation, enhancement, and
zoom, is performed to artificially increase the number of Nexus image dataset
samples to include 20 variants per image.  We then used the trained network to
generate synthetic images: generate a random vector ($z$) drawn from the
uniform distribution, then feed $z$ to DCGAN's generator network to construct
an image. We repeated this process to generate 200,000 images, that form our
{\it synthetic image dataset}. Figure~\ref{fig:trinket}(b) shows sample images
generated by this network, alongside similar images from the Nexus dataset.

\noindent
{\bf Synthetic credential dataset}.
We have generated the binary imageprints for the images in Nexus dataset based
on the best parameters of ai.lock (see $\S$~\ref{sec:eval:parameters}).  For
each considered $\lambda$ value, we consider the value at each position of the
binary imageprint as an independent Bernoulli random variable. We then
calculate the probability of observing a $1$ in each position based
on the imageprints of the Nexus dataset. We use these probabilities to draw
100,000 random samples (of length $\lambda$) from the corresponding Bernoulli
distribution for each position. The resulting random binary imageprints form
our {\it synthetic credential dataset}. We have experimented with 10 values of
$\lambda$ ranging from 50 to 500, thus, this dataset contains 1 million
synthetic imageprints.


\subsection{Evaluation Datasets}
\label{sec:eval:data}

We use the above image datasets to generate {\it authentication samples} that
consist of one candidate image and one reference image.

\noindent
{\bf ai.lock attack dataset}. 
We use roughly 85\% of the images from the Nexus, ALOI and Google datasets 
as a {\it training} set to train and estimate the performance of ai.lock. 
We use the remaining 15\% of images in each dataset (i.e., 220 Nexus, 
3,600 ALOI and 1,178 Google images) as a {\it holdout} set. We use the 
holdout dataset to assess the generalization error of the final model, and 
as a real image attack dataset (see $\S$~\ref{sec:eval:attack}).


We generate the samples in holdout dataset using each subset of Nexus, 
ALOI and Google separately as follows. Each image of the Nexus holdout dataset 
is chosen as a reference image once, then coupled once with all
the other images in the Nexus, ALOI and Google sets, used as candidate images.
Therefore, there are $\frac{220 \times 219}{2} = 24,090$ combinations of
samples for the images in the Nexus set. For each $55$ unique objects in this
set, there are 6 ($4 \choose 2$) possible valid samples that compare one image
of this object to another image of the same object.  Thus, there are
$55 \times 6 = 330$ valid samples in the Nexus set. We then generate
$220 \times 3,600 = 792$K and $220\times1,178 = 259,160$ invalid samples from
comparing Nexus images to images in ALOI and Google sets respectively.
Therefore, the ai.lock holdout set contains a total of $1,075,250$ samples.

In addition, the training set is further divided into $5$ folds, for cross
validation.  Each training fold contains $236$, $4,080$ and $1,335$ images of
Nexus, ALOI and Google datasets respectively.  Therefore, there are $\frac{236
\times 235}{2} = 27,730$ samples for the fold's $59$ unique Nexus set objects,
of which $59\times6 = 354$ pairs are valid. Similarly, we generate
$236\times4,080 = 962,880$ and $236\times1,335 = 315,060$ invalid samples, that
consist of Nexus images coupled with ALOI and Google images, respectively.
Thus, each training fold has a total of $1,305,670$ samples, of which $354$ are
valid.

\noindent
{\bf Synthetic image attack datasets}.
We divide the synthetic image dataset of $\S$~\ref{sec:data}
into 2 equal sets, each containing 100,000 images.  Then, we build two {\it
synthetic image attack datasets} (DS1 and DS2) by repeating the following process 
for each subset of the synthetic image dataset: combine each Nexus dataset image,
used as a reference image, with each image from the subset of the synthetic
image dataset, used as a candidate image. Therefore, in total we have 140 million
samples in each of DS1 and DS2.

\noindent
{\bf Synthetic credential attack dataset}.
We use the synthetic credential dataset described in
$\S$~\ref{sec:data} to build a {\it synthetic credential attack
dataset}: for each value of $\lambda$, combine the imageprint of each Nexus
dataset image, used as a reference imageprint, with each imageprint in
synthetic credential dataset, used as the candidate imageprint. Hence, we have
140 million authentication samples in this dataset for each value of $\lambda$.
We repeat this process for 10 values of $\lambda$, ranging from 50 to 500.
Therefore, in total this dataset contains $10 \times 140$ M = 1.4 billion
samples.

\noindent
{\bf Illumination robustness evaluation dataset}.
To evaluate the performance of ai.lock under illumination changes, we use the
ALOI holdout set ($3,600$ images) that includes up to 11 images of each object
captured under a different illumination condition. Specifically, we pair each
image in the ALOI holdout set (i.e., not used during training) with all the
other images in this set. Therefore, we have a total of $\frac{3600 \times
3599}{2} = 6,478,200$ authentication samples in the illumination robustness
evaluation dataset, of which $6,306$ samples are valid.

\noindent
{\bf Entropy evaluation dataset}.
We randomly selected 2 billion unique pairs of images from the YFCC100M toy
dataset. In each pair, an image is considered to be the reference, the other is
the candidate.



\section{Experimental Evaluation}
\label{sec:eval}

We evaluate ai.lock and its variants. First, we describe the process we used to
identify the best ai.lock parameters. We use these parameters to evaluate the
performance of ai.lock under the attack datasets of $\S$~\ref{sec:eval:data}.
We also show that ai.lock is a $\delta$-LSIM function, empirically estimate its
entropy, and measure its speed on a mobile device. 

\subsection{ai.lock: Parameter Choice}
\label{sec:eval:parameters}

We identify the best parameters for the ai.lock variants
using 5 fold cross validation on the ai.lock training dataset 
(see $\S$~\ref{sec:eval:data}).

\noindent
{\bf Best principal component range}. To identify the best PC range, we use 5
fold cross validation as follows. First, we retrieve the embedding vector
(output of the last hidden layer of Inception.v3) for each image in the ai.lock 
training dataset.
Then, for each cross validation experiments, we use 4 training
folds to find the principal components of the embedding vectors.  Then,
we transform the embedding vectors of the test fold into the newly identified
feature space. Finally, we project them into several randomly generated vectors
(LSH) to construct the binary imageprint of the images. 


We have experimented
with different subsets of the transformed feature space of various size including the
first and second consecutive principal component sets of size 50, 100, 150, and
200, as well as, the first 400 PCs. 

\begin{figure}
\centering
\includegraphics[width=0.49\textwidth]{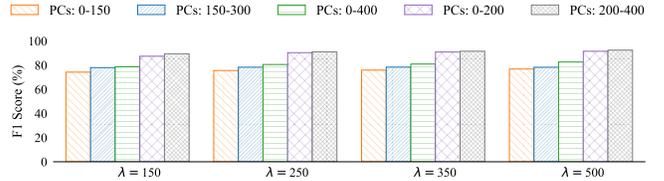}
\caption{Comparison of ai.lock performance (F1 score) when using different
subset of principal component feature ranks for different imageprint length
($\lambda$) values. {\bf PCs ranked 200-400 constantly outperform other tested
subsets.}}
\label{fig:ai:lock:featureranks}
\end{figure}

Figure~\ref{fig:ai:lock:featureranks} shows the cross validation performance
achieved by ai.lock when using different subsets of PC features for different
$\lambda$ values. We observe that the PCs ranked 200-400 perform consistently
the best. This might seem surprising, as higher ranked PCs have higher variability and
thus we expected that they would have more impact in differentiating between
valid and invalid samples. We conjecture that some of the lower rank
coordinates of these transformed vectors are more efficient in capturing the
lower level details of the input object images that differentiate them from other 
object images.

\begin{figure*}[t]
\centering
\includegraphics[width=\textwidth]{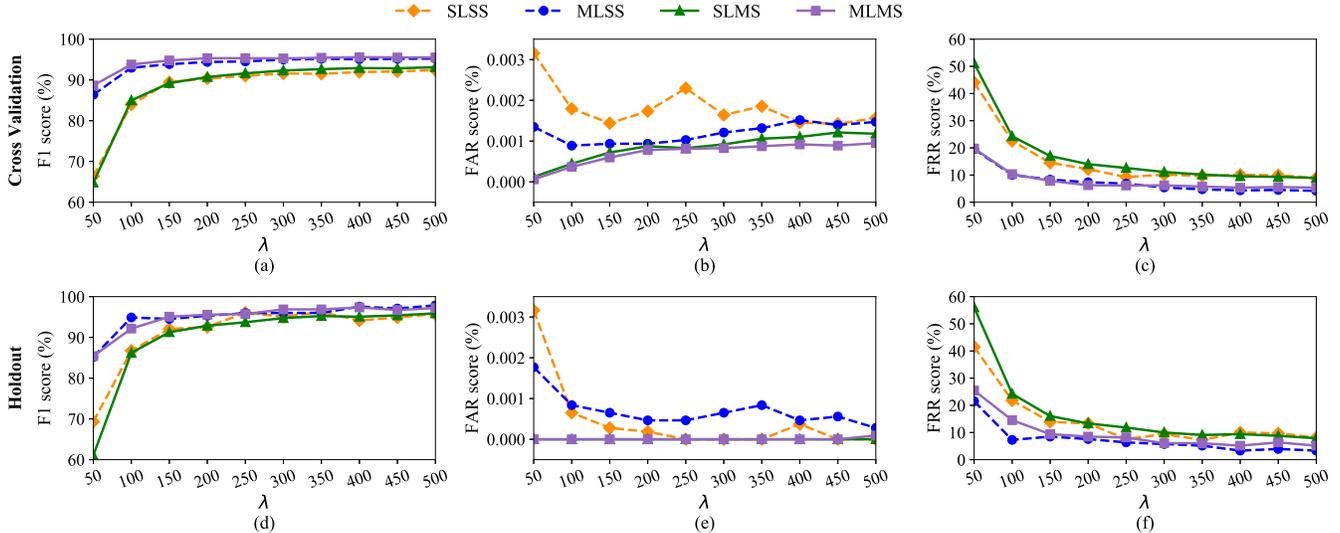}
\caption{(a-c) ai.lock cross validation performance, and (d-f) ai.lock holdout
performance using different ai.lock variants: Single Layer Single Segment
(SLSS), Multi Layer Single Segment (MLSS), Single Layer Multi Segment (SLMS),
Multi Layer Multi Segment (MLMS). {\bf Exploiting information from multiple
Inception.v3 DNN layers (multi layer variants) lowers the FRR, while splitting
images into smaller segments (multi segment variants) lowers the FAR.} The MLMS
variant of ai.lock consistently achieves the lowest FAR, that can be as low as
0\% for the holdout dataset.}
\label{fig:ai.lock:perf:comparison}
\vspace{-10pt}
\end{figure*}


\begin{table}[t]
\centering
\resizebox{0.47\textwidth}{!}{%
\textsf{
\begin{tabular}{c | c c c c c c c c c c}
\toprule
\textbf{$\lambda$}  & \textbf{50} & \textbf{100}& \textbf{150} & \textbf{200} & \textbf{250}& \textbf{300} & \textbf{350}& \textbf{400} & \textbf{450}& \textbf{500}\\
\midrule
$\tau \times 10$ & 7.80 &    7.30 &  7.07 &  6.95 &  6.80 &  6.87 &  6.80 &  6.85 &  6.87 &  6.82\\
\bottomrule
\end{tabular}
}}
\caption{Error tolerance threshold ($\tau$) values for the basic ai.lock
obtained through cross validation over the ai.lock dataset, when using PCs with
feature ranked 200-400.}
\label{table:slss:ett}
\vspace{-15pt}
\end{table}

\noindent
{\bf Identify the best threshold}.
We identify the best threshold that separates the
valid from the invalid authentication samples using the binary imageprints of 
the testing folds in each of the 5 cross validation experiments using ai.lock 
training set.  
Particularly, we normalize the Hamming distance of each pair of
imageprints in the test fold by the length of the imageprints. 
Then, we apply more than $4$K different real values, between 0 and 1, as a 
threshold on the normalized Hamming distances of the authentication pairs 
to classify them. At the end of the 5\ts{th} cross validation experiment, 
we select the threshold that has the maximum average performance, in terms 
of F1 score, as the best separating threshold. We call this the 
{\it Error Tolerance Threshold}, which we denote by $\tau$.

Table \ref{table:slss:ett} reports the $\tau$ values for basic ai.lock with 
different values of $\lambda$, when using PCs with feature rank 200-400. 
We observe that as $\lambda$ increases, the value for $\tau$ decreases: 
we posit that larger $\lambda$ values preserve more information about the 
input vectors (PCs of the embedding vectors) in the LSH output. 

We translate $\tau$ to the error correcting capacity required for ECC.
Specifically, for an imageprint of length $\lambda$, we choose an ECC that is 
able to correct up to $c = \floor{\lambda \times (1-\tau)}$ bits.


\noindent
{\bf ai.lock MLSS variant}.
Similar to the basic ai.lock, we have experimented with multiple ranges of PCs and
$\lambda$ values to identify the $\tau$ values for MLSS ai.lock, using the 5 fold cross
validation experiment on the ai.lock training dataset.



\begin{table}
\centering
\resizebox{0.47\textwidth}{!}{%
\textsf{
\begin{tabular}{c | c c c}
\toprule
$t$ (matching segment counts out of 5) & \textbf{3} & 4 & 5\\
\midrule
F1 score (\%) for SLMS  & \textbf{93.13} & 90.95 & 85.84\\
F1 score (\%) for MLMS  & \textbf{95.53} & 94.64 & 92.42\\
\bottomrule
\end{tabular}
}}
\caption{Cross validation performance (F1 score) for different values of $t$
(number of segments that need to match out of 5) when using PCs with feature
rank 200-400 and $\lambda = 500$ for SLMS and MLMS variants of ai.lock. $t = 3$
consistently achieves the best performance.}
\label{table:cv:t}
\vspace{-15pt}
\end{table}

\noindent
{\bf ai.lock: Multi segment variants}.
For this ai.lock variant, we identify the $\tau$ values separately for each 
image segment, using the 5 cross validation experiment explained above.  
Therefore, we end up having 5 
different $\tau$ values corresponding to each image segment. The $\tau$
corresponding to each segment can be used to identify if there is a match
between the piece of the candidate image to the corresponding piece in the
reference image. We say that the whole candidate and reference images match,
when $t$ of their segments match. We have tested with $t$ ranging from 3 to 5 and
observed that $t$=3 achieved the best F1 score (see Table~\ref{table:cv:t}).

\noindent
{\bf Cross validation performance}.
We now report the cross validation performance of ai.lock with the parameters 
identified above, for $\lambda$ ranging from 50 to 500. 
Figures~\ref{fig:ai.lock:perf:comparison}(a)-(c) compare the F1 score, FAR and
FRR values of the best version of the ai.lock variants (basic SLSS, SLMS, MLSS, and MLMS) 
over the 5-fold cross validation experiments, using ai.lock training
dataset. The performance of all ai.lock variants improves with increasing 
the value of $\lambda$. The MLMS ai.lock achieves the best performance, with
an F1 score of $95.52\%$ and FAR of $0.0009\%$ when $\lambda = 500$.  The MLSS
ai.lock also consistently improves over the basic ai.lock, with a smaller FRR
and a smaller or at most equal FAR.  Its FRR ($4.18\%$ for $\lambda=500$) is
slightly smaller than that of MLMS variants ($5.36\%$), but it exhibits a
slight increase in FAR. For large values of $\lambda$, the FRR of SLMS and SLSS
are almost equivalent. 

The average cross validation Equal Error Rate (EER, the rate at which the FAR =
FRR) of ai.lock for the SLSS and MLSS variants is less than $0.67\%$ and
$0.17\%$ respectively when using PCs with feature rank $200-400$ and $\lambda =
500$.

The purpose of the LSH-based transformation is to encode the feature vector of
an image extracted by a DNN into a binary string. Our conjecture is that larger
lambda values extract more high quality information about the feature vectors,
which in turn leads to lower FAR and FRR. This is partly due to the random
nature of the LSH we used (see Figure~\ref{fig:holdout:hammingsim}), where
roughly half of the bits among different images are different, and images of
the same object have a smaller distance overall. Using more LSH bits reduces
the variance of the distance that was due to perturbations from using a random
projection, hence provides a better separation between TP and FP image
comparisons.

\subsection{Resilience to Illumination Changes}
\label{sec:eval:crossvalidation}


We evaluate the resilience of ai.lock to illumination changes using the
6,478,200 authentication samples of the illumination robustness evaluation
dataset ($\S$~\ref{sec:eval:data}).  While the FAR of the MLMS variant of
ai.lock (for $\lambda=500$ and $t=3$) remains very small ($0.006\%$), its FRR
increases to $16.9\%$.  Decreasing the required matching segments count ($t$)
to 2, reduces the FRR to $11.43\%$, which results in a slightly higher FAR of
$0.010\%$.

\subsection{ai.lock Under Attack}
\label{sec:eval:attack}

\begin{table}
\centering
\footnotesize{
\textsf{
\begin{tabular}{c | c c c c c c c c c c}
\toprule
\textbf{$\lambda$} & \textbf{50} & \textbf{150} & \textbf{250} & \textbf{350} & \textbf{500}\\
\midrule
FAR$\times 10^{+6}$ & 33.87 & 4.34 & 3.29 & 0.69 & 0.20\\
\bottomrule
\end{tabular}
}}
\caption{SLSS ai.lock performance on synthetic attack DS1. The
FAR decreases significantly as $\lambda$ grows from 50 to 500. {\bf The FAR
when $\lambda$ = 500 is only $0.2\times 10^{-6}$}.}
\label{table:singlelayer:singleimage:barbarian}
\vspace{-15pt}
\end{table}

\noindent
{\bf Holdout dataset, real image attack}.
The performance over the ai.lock holdout set is reported in
Figure~\ref{fig:ai.lock:perf:comparison}(d)-(f). As before, the performance of
all the ai.lock variants improves with the increase in $\lambda$.  In agreement
with the results of the cross validation experiments, we conclude that
exploiting information from multiple Inception.v3 layers decreases the FRR,
while using information from multiple image segments decreases the FAR. In
addition, the MLMS ai.lock variant achieves the highest F1 score ($97.21\%$ for
$\lambda = 500$). The SLMS and MLMS schema consistently
achieve the lowest FAR, which is as low as 0\% on the holdout dataset.

\noindent
{\bf Synthetic image attack}.
We use the synthetic attack dataset DS1 of
$\S$~\ref{sec:eval:data} to evaluate the performance of SLSS ai.lock,
using the trained parameters of $\S$~\ref{sec:eval:parameters}. 
%
%
%
Table~\ref{table:singlelayer:singleimage:barbarian} shows the performance of
ai.lock in classifying these attack samples. The FAR decreases significantly
with $\lambda$, and is as low as 0.00002\% when $\lambda$ = 500.

The proportion of the reference images that have been broken at least once 
decreases significantly by increasing $\lambda$: from $16.86\%$ to
0.79\% (11 Nexus images) when $\lambda$ is $150$ and $500$ respectively.  
A majority of the broken references are broken only by a small number of 
candidate images:
%
%
when $\lambda=500$, only 2 of the 11 broken images have been broken 5 times by 
the synthetic images in DS1. The average number of trials until finding the first 
matching synthetic image, over the 11 broken reference images, is 31,800.


\begin{figure}
\centering
\includegraphics[width=0.49\textwidth]{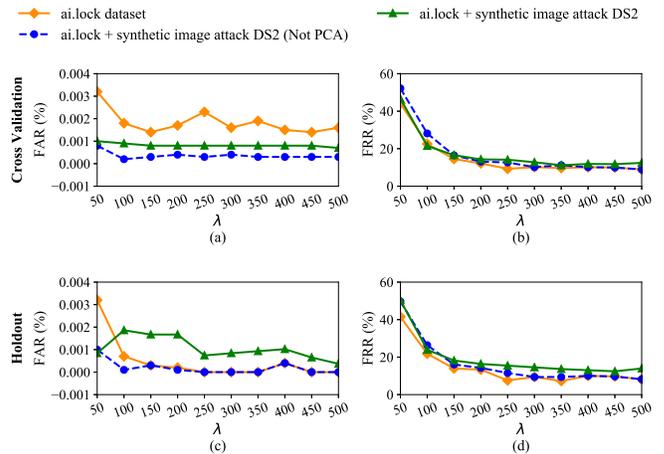}
\caption{(a) Cross validation FAR, (b) Cross validation FRR , (c) Holdout FAR,
and (d) Holdout FRR of SLSS ai.lock when trained over the ai.lock and synthetic
image attacks of DS2.}
\label{fig:train:barbarian}
\vspace{-12pt}
\end{figure}

\noindent
{\bf Vaccinated ai.lock}.
To further improve the ai.lock resistance to synthetic image 
attacks, we use the synthetic image attack dataset DS2 (see
$\S$~\ref{sec:eval:data}) along with the ai.lock training dataset, to train
ai.lock. Specifically, we divide the synthetic image attack dataset DS2 into 5
folds and distribute them into the 5 training folds of the ai.lock dataset.  In
other words, we train ai.lock on an additional 236 $\times$ 20,000 = 4,720,000
invalid authentication samples. The holdout set remains untouched and is used
to evaluate the effectiveness of this approach. Then, we train ai.lock with SLSS 
as before using the cross validation experiment (see $\S$~\ref{sec:eval:parameters}).

We experimented with two cases. First, the invalid synthetic image attack
samples in DS2 contribute to both the PCA based feature selection and the 
error tolerant threshold 
($\tau$) discovery processes. Second, those samples are only used in the
process of discovering $\tau$. Figure~\ref{fig:train:barbarian} shows the cross
validation FAR and FRR (a, b) as well as the performance over the holdout set
(c, d). In both experiments, we observed a drop in the FAR of ai.lock, however,
the FRR increases. The FAR improvement is higher for the second case. We
conjecture that the inclusion of synthetic, not camera captured images, is
misleading the PCA based feature selection module into capturing irrelevant
information.

\begin{figure}
\centering
\includegraphics[width=0.49\textwidth,keepaspectratio]{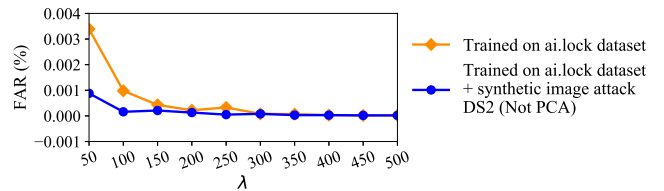}
\caption{FAR of ai.lock on synthetic image attack, when trained on the ai.lock
dataset vs. when trained also on DS2. {\bf The ``vaccinated'' ai.lock 
improves its resistance to the synthetic image attack}: the FAR
drops by more than $74\%$, $51\%$ and $59\%$ when $\lambda$ is $50$, $150$ and
$350$ respectively.}
\label{fig:train:barbarian:performance}
\vspace{-10pt}
\end{figure}

We used the ai.lock trained on the synthetic image attack dataset DS2 to
evaluate its performance over the synthetic image attack DS1.
Figure~\ref{fig:train:barbarian:performance} compares the performance of
ai.lock when trained on the ai.lock dataset and when trained on the ai.lock and
the synthetic dataset DS2.  Training also over synthetic image attack samples
helps ai.lock to be more resilient to synthetic image attack, 
especially for small values of $\lambda$. 

\begin{table}
\centering
\footnotesize{
\textsf{
\begin{tabular}{c | c c c c c c c c c c}
\toprule
\textbf{$\lambda$} & \textbf{50} & \textbf{150} & \textbf{250} & \textbf{350} & \textbf{500}\\
\midrule
FAR$\times 10^{+6}$ & 11.89 & 0.09 & 0.03 & 0.000 & 0.000\\
\bottomrule
\end{tabular}
}}
\caption{SLSS ai.lock performance on the synthetic credential attack. {\bf
ai.lock is unbreakable under 1.4 billion samples of the synthetic credential
attack}: its FAR is 0 when $\lambda \ge$ 300.}
\label{table:singlelayer:singleimage:randombinary}
\vspace{-15pt}
\end{table}

\noindent
{\bf Synthetic credential attack}.
Table~\ref{table:singlelayer:singleimage:randombinary} shows the FAR values
for ai.lock under the synthetic credential attack dataset described in
$\S$~\ref{sec:eval:data}.
%
%
For all values of $\lambda$ greater than 300, the FAR of ai.lock is equal to
0. Even for a $\lambda$ of 50, the FAR is $11.89 \times 10^{-4}$\%.  This
is an important result: even a powerful adversary who can create
and test synthetic credentials on a large scale, is unable to break the ai.lock
authentication.

\subsection{Is ai.lock $\delta$-LSIM?}
\label{sec:eval:lsim}

We now evaluate if the basic ai.lock (SLSS) variant, with the parameters
identified in $\S$~\ref{sec:eval:parameters} preserves the similarity of the
input space, i.e., if it satisfies the LSIM properties (see
Definition~\ref{def:dlsim}). We use the ai.lock holdout set to evaluate the
probability of obtaining the same hash value for valid and invalid samples.

Let $\pi_i$ and $\pi_j$ be the imageprints corresponding to two images in the
ai.lock holdout set. Let $d_H(\pi_i,\pi_j)$ denote the Hamming distance and $S_H(\pi_i,\pi_j)$ denote the normalized {\it Hamming similarity} of
these imageprints, i.e., $S_H(\pi_i,\pi_j) = 1 - \frac{d_H(\pi_i,\pi_j)}{\lambda}$.

The output of ai.lock can be considered either as a single bit or a string of
bits. In the former case, the imageprints consist of the concatenation of the
output of multiple hash functions, while in the later case, the entire
imageprint is assumed to be the ai.lock hash value. In the following, we
empirically evaluate the $P_1$ and $P_2$ values (see Definition $\S$\ref{def:dlsim}),
for the case where the entire ai.lock imageprint is considered as the hash value. 
In Appendix~\ref{sec:appendix:lsim}, we further show that ai.lock is also a
$\delta$-LSIM function when its hash value is a single bit.

\begin{table}
\centering
\footnotesize{
\textsf{
\begin{tabular}{c | c c c}
\toprule
\textbf{$\lambda$} & \textbf{150} & \textbf{350} & \textbf{500}\\
\midrule
$P_1$ & 8.6e-1 & 9.3e-1 & 9.1e-1\\
$P_2$ & 2.8e-6 & 0.0 & 0.0\\
\bottomrule
\end{tabular}
}}

\caption{
Average probability of collision, for valid ($P_1$) and invalid
($P_2$) samples in the ai.lock holdout set, when the ai.lock 
imageprint is considered as image hash value and at
most $c = \floor{\lambda \times (1-\tau)}$ bits of error is allowed. 
{\bf In all cases, $P_1 > P_2$, thus conclude that ai.lock is an LSIM function}.}
\label{table:holdout:avg:collision:prob:multi}
\vspace{-15pt}
\end{table}

We set $\delta = \tau$, where $\tau$ is the error tolerance
threshold obtained from the ai.lock training process (see Table~\ref{table:slss:ett}), 
for different values of $\lambda$.
Table~\ref{table:holdout:avg:collision:prob:multi} shows the $P_1$ and $P_2$
values achieved by the basic ai.lock over the holdout dataset. 
We perform Mann-Whitney one-sided test with
alternative hypothesis $P1>P2$. Based on the observed $p-value = 0.00$, 
($\alpha=0.05$), for different values of $\lambda$, we conclude that the alternative 
hypothesis is true, hence, ai.lock is a $\delta$-LSIM function over the holdout dataset.

\subsection{On the Entropy of Imageprints}
\label{sec:eval:entropy}

We have used the entropy evaluation dataset (see $\S$~\ref{sec:eval:data}) to 
empirically calculate the entropy of the
imageprints generated by the ai.lock variants.  The empirical entropy of an
authentication solution is propositional to the size of the keyspace that the
attacker needs to search to find a match for the authentication secret. For
biometric information, estimating this size is difficult.  In such cases, the
entropy can be estimated as
$-log_2(\frac{1}{FAR})$~\cite{entropyBasedOnFAR}.  We performed this study for
different values of $\lambda$ and the best parameter choice of ai.lock 
(see $\S$~\ref{sec:eval:parameters}), using the entropy evaluation dataset. 

On the 2 billion image pairs in the entropy evaluation dataset, 
the FAR of the SLSS ai.lock variant is $0.020\%$ and $0.035\%$ when $\lambda$
is  50 and 500 respectively, for an entropy of 12.28 bits and 11.48 bits. We
have visually inspected several hundreds of image pairs that resulted in false
accepts and observed that a significant proportion were due to images that
contained the same object type, e.g. ribbons, helmets, etc.  This result is not
unexpected: the SLSS variant uses only the last hidden layer of 
Inception.v3 network. Since Inception.v3 is trained for image classification task,
it is expected to have similar activations on the last hidden layer for images
of the same object type. We expect to eliminate this situation by requiring the  
match between activations of multiple inception layers (multi layer variant). 

The FAR of the MLMS ai.lock variant on the entropy evaluation dataset, for
$\lambda$ values of 500 and 150, is $0.0007\%$ and $0.0004\%$ respectively.
Therefore, the estimated entropy of ai.lock imageprints is $17.14$
and $18.02$ bits respectively.

%

\subsection{ai.lock Speed}
\label{sec:eval:speed}

\begin{table}
\centering
\resizebox{0.47\textwidth}{!}{%
\textsf{
\begin{tabular}{c | c c c c c c c c c c}
\toprule
\textbf{$\lambda$} & \textbf{150} & \textbf{250}& \textbf{350} &  \textbf{500}\\
\midrule
DI2E module (Inception v.1) & 0.7 & 0.7 & 0.7 & 0.7\\
DI2E module (Inception v.3) & 1.9 & 1.9 & 1.9 & 1.9\\
PCA + LSH module  & 0.044 & 0.049 & 0.051 & 0.066\\
\bottomrule
\end{tabular}
}}
\caption{Processing time (in seconds) of SLSS ai.lock modules, for
different values of $\lambda$.
When using Inception.h5, {\bf the overall ai.lock speed is below 0.8s.}}
\label{table:ai.lock:speed}
\vspace{-20pt}
\end{table}

We have implemented ai.lock using Android 7.1.1 and Tensorflow 0.12.1 and have
evaluated its speed using 1,000 images of the Nexus dataset on a Nexus 6P
smartphone (Qualcomm Snapdragon 810 CPU and 3GB RAM).
Table~\ref{table:ai.lock:speed} shows the average processing time of the 3 main
ai.lock modules for different values of $\lambda$.
Independent of the value for $\lambda$, ai.lock's DI2E module takes 1.9s to compute 
the activations of all the layers of Inception.v3. 
When using Inception.h5~\cite{inceptionh5} (a smaller network), DI2E module takes 0.7s.
The combined PCA and LSH speed increases with the value of $\lambda$, but is
below 70ms for $\lambda$ = 500.
The processing overhead of ai.lock is below 2s and 1s using Inception.v3 
and Inception.h5 respectively.

To minimize its impact on user experience on a Nexus 6P, ai.lock needs to use
Inception.h5. The most significant processing overhead of ai.lock
is on computing the activation of the DNN, which directly depends on the size
of the network. Note that compressing the network using the DNN distillation
approach~\cite{distilling} can alleviate this overhead. Future device and
Inception improvements will likely improve the ai.lock performance and
accuracy.

\section{Discussion and Limitations}
\label{sec:discussion}

\noindent
{\bf Default authentication, revocation and recovery}.
If the image based authentication fails a number of times or the ai.lock secret 
is not available, the authentication falls back to the default authentication 
mechanism, e.g. text passwords.

\noindent
{\bf Strong passwords}.
ai.lock benefits from users choosing strong, high-entropy and unique objects
for authentication. ai.lock can use datasets of images of frequently occurring, thus low entropy, objects and learn to reject similar objects during their registration by the
user. Further, the image classification task can be adapted to detect images
belonging to classes of weak, low-entropy authentication objects. In addition,
similar to text passwords, users could be encouraged to pick an ordered
combination of personal objects for authentication.



\noindent
{\bf Usability}.
Although usability is not the focus of this paper, we expect ai.lock to share
several limitations with face based authentication mechanisms due to their
similarities in the form factor. These include susceptibility to inappropriate
lighting conditions~\cite{2015biometriciphoneandroid}.  While the FAR of
ai.lock remains small under illumination changes, its FRR increases, affecting
its usability.  However, DNNs are capable of learning representations that are
invariant to input changes, e.g. lighting, translation, etc. Thus, the DI2E
module of ai.lock can be further fine-tuned to be more resistant to
illumination changes. We leave the investigation of such improvement for future
work.

In~\cite{ATC17} we have evaluated the usability aspects of an image based
authentication approach, and have shown that (1) the user entry time was
significantly shorter compared to text passwords on a mobile device, (2) the
participants were able to remember their authentication objects 2 and 7 days
after registering them, and (3) the participants perceived object based
authentication to be easier to use than text passwords, and were willing to
adopt it. Further studies are required to understand (1) the user choice of the
secret objects or scenes and whether it impacts the secret key space, (2) the
ability of ai.lock to filter out common or low-entropy images, (3) the
scenarios where users are willing to adopt ai.lock authentication and (4) other
limitations associated to ai.lock authentication.

\noindent
{\bf Shoulder surfing}.
Similar to face based authentication, ai.lock is vulnerable to shoulder surfing
attacks where the adversary captures images of the objects or scenes used by
victims.  However, ai.lock eliminates remote attacks,
e.g.,~\cite{facebookface}, moves the target away from sensitive body features,
and enables users to trivially change their image-passwords. Similar to
biometrics, ai.lock can also benefit from liveness verification
techniques~\cite{RTC17}, that ensure that the adversary has physical access to
the authentication object or scene, to prevent sophisticated image replay
attacks. In addition, in Appendix~\ref{sec:appendix:shouldersurfing} we show
that the knowledge of the authentication object type does not provide the
adversary with significant advantage when launching a brute force attack.

\noindent
{\bf Multi-factor authentication}.
ai.lock can also be used in conjunction with other authentication
solutions. For instance, the image password set and authentication steps
described in $\S$~\ref{sec:background:ss} can take advantage of a secondary
secret (e.g. password, PIN), increasing the number of authentication factors to
improve security. To this end, let $r$ be a random salt.  We modify $x$ in the
fuzzy biometric protection solution outlined in $\S$~\ref{sec:background:ss} to
be the randomized hash of the secondary secret computed using salt $r$.
Randomized hashing ensures the required formatting and bit length for $x$ 
can be achieved using key derivation function (e.g. HKDF ~\cite{HKDF}), etc.
The random salt $r$ needs to be stored along with the other authentication
credentials, i.e. $SS(R,x)$.


\noindent
{\bf Compromised device}.
Our model assumes an adversary that physically captures a victim’s device and
thus has black-box access to the authentication function. ai.lock is not
resilient to an adversary who installs malware on the victim device. Such
malware may for instance leverage PlaceRaider~\cite{TRCK13} to construct
three dimensional models of the environment surrounding the victim, including
of the authentication object.

Trusted hardware can secure ai.lock and even obviate the need for
secure sketches. However, it would reduce the number of devices where
ai.lock can be applied. Techniques similar to AuDroid~\cite{PSJA15} could be
employed to ensure that unauthorized processes or external parties cannot
access and misuse the device camera, however, they may still leave ai.lock
vulnerable to cache attacks~\cite{LGSM16}.

\section{Conclusions}

In this paper, we introduced ai.lock, the first secure and efficient image based
authentication system. We
have presented a suite of practical yet powerful image based attacks and built
large scale attack datasets. We have shown that even under our powerful
attacks, ai.lock achieves better entropy than state-of-the-art biometric
authentication solutions.


%

\section{Acknowledgments}

We thank the shepherd and reviewers for their excellent feedback. This research
was supported in part by grants from the NSF (CNS-1526494, CNS-1527153 and
SES-1450619) and the Florida Center for Cybersecurity.

\bibliographystyle{ACM-Reference-Format}
\bibliography{biometric,bogdan,dnn,entropy,image,lsh,privacy,vault,application,authentication,hci}

\appendix

\section{Motivation for feature selection using PCA}
\label{sec:appendix:pca}

\begin{figure}[H]
\centering
\includegraphics[width=0.49\textwidth]{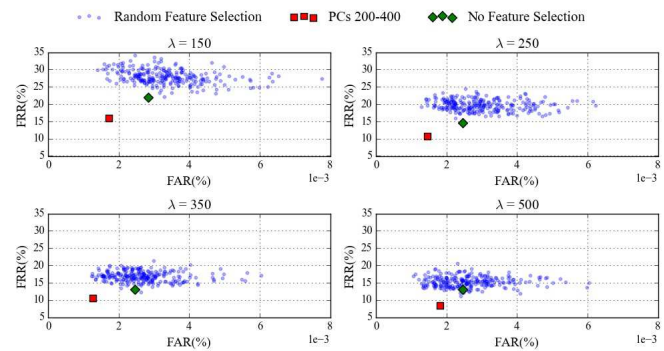}
\caption{{\bf PCA motivation}: FRR vs. FAR of (i) ai.lock when using PCA (with
features ranked 200-400), (ii) ai.lock with no feature selection (``Raw''), and
(iii) 250 independent instances of ai.lock when using a feature selection
approach that randomly selects 200 features. {\bf ai.lock with PCA consistently
achieves the lowest FRR and often the lowest FAR}. 
}
\label{fig:ai:lock:pca}
\vspace{-10pt}
\end{figure}

We now justify the need for the PCA step of ai.lock. For this, we compare the
best version of ai.lock running PCA (i.e., features ranked 200-400), with two
other versions. First, we consider a baseline version (which we call ``Raw''),
that uses no feature selection component. Specifically, Raw applies LSH to the
raw embedding vectors, then, identifies the best threshold $\tau$ using the
5-fold cross validation experiment described in $\S$~\ref{sec:eval:parameters} for ai.lock.
Second, we compare against an ai.lock variant where we replace
the PCA component with a random choice of 200 features (of the embedding vectors) produced by
the last hidden layer of Inception.v3. Figure~\ref{fig:ai:lock:pca} shows the results
of this comparison for $\lambda$ values of 150, 250, 350 and 500, and 250
different instances of ai.lock with random feature selection.  We observe that
ai.lock with PCs of rank 200-400 consistently achieves the significantly lower FRR, 
and often the lowest FAR. In addition, we observe that randomly choosing the features is
not ideal, as it often performs worse than when no feature selection is used at
all.

\section{Object/Scene Guessing Attack}
\label{sec:appendix:shouldersurfing}

\begin{table}
\centering
\resizebox{0.49\textwidth}{!}{%
\footnotesize{
\textsf{
\begin{tabular}{c | c c c c}
\toprule
\textbf{\# of words in image search query} & \textbf{1} & \textbf{2} & \textbf{3} & \textbf{4}\\
\midrule
Dataset size ($d_{size})$ & $12,413$ & $24,882$ & $26,418$ & $26,766$ \\
Avg \# of trials before FA (random order) & $12,078$ & $23,205$ & $24,641$ & $25,028$\\
Avg \#  of trials before FA (guessing attack) & $12,034$ & $22,755$ & $23,921$ & $24,488$\\
\midrule
Portion of broken references (\%) & $5.0$ & $9.0$ & $10.9$ & $9.0$ \\
\bottomrule
\end{tabular}
}}}
\caption{ai.lock under the object guessing attack.  The average number of
trials before the first false accept (FA) drops only slightly in the object
guessing attack scenario when compared to a random ordering of attack images.
Thus, knowledge of the authentication object type provides the adversary only
nominal guessing advantage.}
\label{table:slss:shouldersurfing}
\vspace{-15pt}
\end{table}

\noindent
{\bf Data}.
We have asked a graduate student to tag each of the 55 unique object images in
the Nexus holdout set with 1 to 4 words. For each value of the number of tags
per image (i.e., 1 to 4), and each object image, we collected 300-500 images
provided by Google's image search engine. Thus, we generated 4 Google image
datasets, one for images found when searching with 1 tag, another when
searching with 2 tags, etc. In total, we have collected 90,479 images.

\noindent
{\bf ai.lock performance under object guessing attack}.
We use the 4 collected image datasets from Google to generate a total of 
$19,905,380$ ``guessing attack'' authentication samples, and use them to
evaluate the {\it guessing entropy}~\cite{DMR04} of ai.lock under an
object/scene guessing attack (see $\S$~\ref{sec:model}). 

Specifically, using each of the 4 Google image datasets we perform the following
two brute force attacks. The first attack emulates an object guessing attack:
re-order the images in the Google dataset to start the brute force attack with
the images of the same object type, then continue with images of other object
categories in a random order. Finally, count the number of trials before the
first match (false accept) occurs.  The second attack is a standard brute force
attack: randomly shuffle the images in the Google image dataset and use them to
brute force each image in the Nexus holdout set. We use the second attack as a
baseline, to determine if knowledge of the object type impacts the trial count
to success.  In both attacks, we count each of the unbreakable reference images
as ``success'' at $d_{size}$ trials, where $d_{size}$ is the number of
images in the corresponding Google image dataset (see Table~\ref{table:slss:shouldersurfing}).

Table~\ref{table:slss:shouldersurfing} summarizes the ai.lock performance under
the object/scene guessing attack scenario. We observe an increase in the
portion of the Nexus images that are broken when the simulated adversary uses
more words to describe the authentication objects for collecting the attack
image dataset.  However, for all experiments, the average number of trials
before success drops only slightly in the object guessing attack scenario
compared to the baseline. This is due to the fact that the reference images
were mostly broken with images of different object categories. We conclude that
knowledge of the secret object type does not provide the adversary with a
significant guessing advantage.

\section{Is ai.lock $\delta$-LSIM for ai.lock with single bit hash value?}
\label{sec:appendix:lsim}

We now show that ai.lock with a single bit hash value is a $\delta$-LSIM 
(see Definition \ref{def:dlsim}). 

ai.lock uses Charikar's random projection LSH ~\cite{C02}. Therefore, 
for any embedding vector (the input to LSH function) 
$u$ and $v$, $Pr[1\ bit\ collision] = 1 - \frac{\theta(u,v)}{\pi}$, 
where $\theta(u,v)$ denotes the angle between
$u$ and $v$.  We use the angle between the feature vectors of images in the
ai.lock holdout set to compute the average probability of collision: 0.79 for
valid samples and 0.50 for invalid authentication samples.

\begin{figure}
\centering
\includegraphics[width=0.49\textwidth]{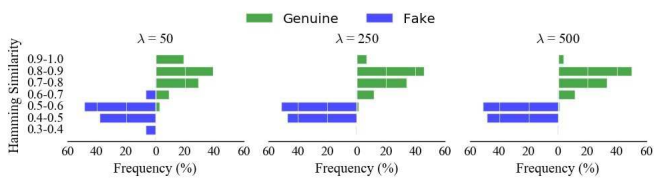}
\caption{Histograms of normalized Hamming similarity between imageprints of valid and
invalid authentication samples in the ai.lock holdout set. The red rectangles
pinpoint the focus areas: valid samples with Hamming similarity below 0.6 and
invalid samples with Hamming similarity above 0.6. {\bf Higher values of $\lambda$ 
provide more effective separation between valid and invalid samples}: when
$\lambda$ = 500, no invalid samples have similarity above 0.6.
}
\label{fig:holdout:hammingsim}
\vspace{-5pt}
\end{figure}


\begin{table}
\centering
\footnotesize{
\textsf{
\begin{tabular}{c | c c c}
\toprule
\textbf{$\lambda$} & \textbf{150} & \textbf{350} & \textbf{500}\\
\midrule
$P_1$ & 0.799 & 0.797 & 0.796\\
$P_2$ & 0.500 & 0.500 & 0.500\\
\bottomrule
\end{tabular}
}}
\caption{
Average probability of collision, for valid ($P_1$) and invalid 
($P_2$) samples in the ai.lock holdout set per imageprint bit basis.
{\bf In all cases, $P_1 > P_2$, thus conclude that ai.lock with 
single bit hash value is an LSIM function}.}
\label{table:holdout:avg:collision:prob:single}
\vspace{-15pt}
\end{table}

Figure~\ref{fig:holdout:hammingsim} shows the histogram of normalized Hamming similarity
between imageprints in the valid and invalid samples of the ai.lock holdout
set.  Unsurprisingly, most invalid samples have a Hamming similarity between
0.4 and 0.6: different images have imageprints that are similar in around half
of their bits (see also Table~\ref{table:holdout:avg:collision:prob:single}).  We
observe that the overlap between the Hamming similarities of valid and invalid
samples significantly reduces for higher values of $\lambda$.

In addition, we compute these probabilities empirically by counting the number
of times when the hash values collide for valid and invalid samples, after the
LSH transformation. We then use this count to compute the average probability
of collision for a valid ($P_1$) and invalid ($P_2$) authentication samples 
(see Table~\ref{table:holdout:avg:collision:prob:single}). We observe the
remarkable similarity of these values, to the ones above, computed
analytically. As $\lambda$ increases, the empirical $P_1$ approaches the
analytic lower bound (0.79). We perform a Mann-Whitney one-sided test with
alternative hypothesis $P1>P2$. 
This test suggests that there is a significant gap between $P_1$ and $P_2$ 
($p-value = 0.00, \alpha=0.05$) for all cases, 
hence, ai.lock is a $\delta$-LSIM on the Nexus holdout dataset.

In addition, comparing the values for $P1$ and $P2$ with the ones reported in 
$\S$~\ref{sec:eval:lsim} for ai.lock with multi-bit hash value, we observe 
that concatenating multiple hashes enlarges the gap between $P_1$ and $P_2$ 
values.




\end{document}